%% file: manuscript.tex
\documentclass[a4paper,fleqn,useAMS,usenatbib]{mnras}

\pdfoutput=1

\usepackage{mathptmx}
\usepackage[T1]{fontenc}
\usepackage{ae,aecompl}

\usepackage[dvips]{graphicx}
\usepackage{amsmath}
\usepackage{amsfonts}
\usepackage{amssymb}
\usepackage{comment}
\usepackage{pdflscape}
\usepackage[dvipsnames]{xcolor}
\usepackage[%
        ]{hyperref}
        
\hypersetup{
	colorlinks=true,	% false: boxed links; true: colored links
	urlcolor=MidnightBlue,		% color of external links
	pdfpagelabels=true,
	hypertexnames=true,
	plainpages=false,
	naturalnames=true,
	pdftitle={Forecasted masses for seven thousand KOIs]},    % title
	pdfauthor={Chen \& Kipping},     % author
	linkcolor=WildStrawberry,          % color of internal links (change box color with linkbordercolor)
	citecolor=ForestGreen,        % color of links to bibliography
}

\input{shortcuts.tex}

\title[Forecasted masses for seven thousand KOIs]{Forecasted masses for seven thousand KOIs}
\author[Chen \& Kipping]{Jingjing Chen$^{1}$\thanks{E-mail:
\href{mailto:jchen@astro.columbia.edu}{jchen@astro.columbia.edu}} \& David M. Kipping$^{1}$\\
$^{1}$Dept. of Astronomy, Columbia University, 550 W 120th Street, New York NY 10027}

\date{Accepted . Received ; in original form }

\pubyear{2017}

\begin{document}
\label{firstpage}
\pagerange{\pageref{firstpage}--\pageref{lastpage}}
\maketitle

\begin{abstract}

Recent transit surveys have discovered thousands of planetary candidates with
directly measured radii, but only a small fraction have measured masses.
Planetary mass is crucial in assessing the feasibility of numerous
observational signatures, such as radial velocities (RVs), atmospheres, moons
and rings. In the absence of a direct measurement, a data-driven, probabilistic
forecast enables observational planning and so here we compute posterior
distributions for the forecasted mass of approximately seven thousand
\textit{Kepler} Objects of Interest (KOIs). Our forecasts reveal that the
predicted RV amplitudes of Neptunian planets are relatively consistent,
as a result of transit survey detection bias, hovering around the few m/s level.
We find that mass forecasts are unlikely to improve through more precise
planetary radii, with the error budget presently dominated by the intrinsic
model uncertainty. Our forecasts identify a couple of dozen KOIs near the
Terran-Neptunian divide with particularly large RV semi-amplitudes which could
be promising targets to follow-up, particularly in the near-IR. With several
more transit surveys planned in the near-future, the need to quickly forecast
observational signatures is likely to grow and the work here provides a
template example of such calculations.

\end{abstract}


\section{Introduction}
\label{sec:intro}

Two of the most fundamental properties of any planet are its mass and radius.
Unfortunately, for the vast majority of the thousands of planetary candidates
discovered by \textit{Kepler} \citep{coughlin:2016}, we only have measurements
of planetary radii. Whilst a direct measurement of mass is always preferable,
predicting a planet's mass based off its radius is often useful. Specifically,
there are numerous follow-up observations for which the predicted
signal-to-noise depends strongly upon the planetary mass.

This is particularly important because \textit{Kepler} has delivered so many
planetary candidates that it is often impractical to schedule follow-up of
every object. Finite resources demand prioritization and one obvious criteria
for ranking the objects is whether an observational signature is even expected
to be detectable.

We highlight several effects where planetary mass directly controls the
amplitude and/or feasibility, such as radial velocity semi-amplitude
\citep{struve:1952}, astrometric amplitude \citep{jacob:1855}, transit
spectroscopy scale heights \citep{seager:2000}, Doppler beaming
\citep{rybicki:1979}, ellipsoidal variations \citep{kopal:1959}, transit
timing variations \citep{holman:2005,agol:2005}, stability of
rings \citep{schlichting:2011}, stability of exomoons \citep{barnes:2002}
and detectability of exomoons \citep{kipping:2009a,kipping:2009b}.

In each case, it is generally preferable to estimate a credible interval for
the planetary mass, rather than a point estimate. Such a credible interval
describes the probable range (at some chosen level of confidence), ideally
accounting for both the present measurement error on the planetary radius and
also the inherent uncertainty caused by using what, practically speaking, will
always be an imperfect predictive model. In statistical parlance,
what we are really describing here is generating posterior samples for the
predicted mass using a probabilistic model conditioned upon posterior samples
of the observed planetary radius. In this way, we explicitly acknowledge
that our predictions are conditioned not just upon a measured radius with
finite uncertainty, but also upon a model which too has finite uncertainty.

Probabilistic forecasts made in this way allow for more robust and
accurate predictions, albeit at the expense of larger credible intervals,
as first highlighted by \citet{wolfgang:2016}.
Whilst it has become increasingly common practice for the exoplanet community
to share posterior distributions of measured parameters such as
planetary radii (e.g. \citealt{rowe:2015,dfm:2016,jupiter:2017}), there remains
a paucity of probabilistic models able to convert these measurements into
a planetary mass, with most mass-radius relations still relying on deterministic
formalisms (e.g. recent examples \citealt{weiss:2014,hatzes:2015,millis:2017}).

The recent availability of a homogeneous set of joint posteriors
for \textit{Kepler} host stars \citep{mathur:2017} and associated transit
parameters \citep{rowe:2015}, combined with the first probabilistic forecasting
mass-radius relation spanning the entire planetary regime \citep{chen:2017},
finally enables mass forecasts for thousands of planetary candidates. In this
work, we forecast the mass of approximately seven thousand \textit{Kepler}
Objects of Interest (KOIs), described in detail in
Section~\ref{sec:forecasting}. We highlight some important implications
and patterns evident from our analysis in Section~\ref{sec:implications},
before framing our results in a broader context in
Section~\ref{sec:discussion}.

\section{Forecasting KOI Masses}
\label{sec:forecasting}

\input{forecasting.tex}

\section{Implications \& Limitations}
\label{sec:implications}

\input{implications.tex}

\section{Discussion}
\label{sec:discussion}

In this work, we have used a data-driven and probabilistic mass-radius relation
to forecast the masses of approximately seven thousands KOIs with careful
attention to correctly propagate parameter covariances and uncertainties
(see \citealt{wolfgang:2016} and \citealt{chen:2017}). We
report 1 and 2\,$\sigma$ credible intervals for each KOI in
Table~2, including derived parameters such as density, surface
gravity and radial velocity semi-amplitude. Full joint posterior samples are
made publicly available at \httplink (https://github.com/chenjj2/forecasts). We stress that these results should be
treated as informed, probabilistic and model-conditioned \textit{predictions}
and not \textit{measurements}.

\forecaster\ has already seen numerous applications (e.g. see
\citealt{dfm:201607,ober:2016, dressing:2017,rod:2017}) to individual systems
or small subsets, but we believe this to be the first application to a large
(several hundred) ensemble of planets ensuring homogeneous methodology.
This work serves as a demonstration of ensemble predictions for planetary
masses based off transit-derived radii. In the near-future, one can expect the
number of planetary candidates discovered through transits to continue to
rapidly grow, exacerbating the demands on follow-up resources, for which
planetary mass is frequently a key driver for various observational
signatures (e.g. radial velocities, atmospheric scale heights, ring/moon
stability). For example, \textit{TESS} is expected to discover $\sim5,000$
transiting planets \citep{bouma:2017}, LSST a further
$\sim\mathcal{O}[10^3]$-$\mathcal{O}[10^4]$ \citep{lund:2015} and
$\sim\mathcal{O}[10^4]$ from PLATO \citep{rauer:2014}. In choosing the subset
of targets for follow-up, forecasted masses will frequently be beneficial, and
thus the analysis provided in this work can be considered as a template for
these future surveys too.

As an example, we have shown how this approach can quickly identify small KOIs
for which the radial velocity amplitudes are unusually high, increasing the
chances for detection (see Section~\ref{sub:opportunities}). The combination of
their forecasted mass, low parent star mass and short orbital period maximizes
$K$. However, the deep-sky nature of \textit{Kepler}, and the bias towards
low-mass stars, means these targets are faint - a problem likely to be resolved
with the all-sky surveys of \textit{TESS} and PLATO. Of course, realistic
target selection will depend upon more than just signal amplitude, likely
factoring in expected noise, planet properties and other program-specific
questions of interest (e.g. see \citealt{lam:2017} for an example of
target selection based on forecasted multiplicity).

Future improvements to the forecasts made in this work are unlikely to result
from simply obtaining more precise planetary radii, since the predicted mass
uncertainties are observed to be dominated by the intrinsic dispersion inherent
to the model itself (see Section~\ref{sub:patterns}). Future work could attempt
to build an updated probabilistic model using more data and likely a second
dependent variable, such as insolation for example. Until that time, the forecasts 
made here are likely the most credible estimates available without direct observation.

\section*{Acknowledgments}

Thanks to Ruth Angus for helpful suggestions in preparing this manuscript.

%\appendix
%
%\input{lambda.tex}

\bsp
\label{lastpage}
\end{document}

%% file: shortcuts.tex
\newcommand{\forecaster}{{\tt forecaster}}

\newcommand{\actuallink}{https://github.com/chenjj2/forecasts}
\newcommand{\httplink}{\href{\actuallink}{this URL}}

\newcommand{\orob} {R}
\newcommand{\re}{R_{\oplus}}
\newcommand{\powerconst}{C}
\newcommand{\omob} {M}
\newcommand{\me}{M_{\oplus}}
\newcommand{\slope} {S}
\newcommand{\logm} {\mathcal{M}}
\newcommand{\logr} {\mathcal{R}}
\newcommand{\logmeq} {\log_{10}(\omob/\me)}
\newcommand{\logreq} {\log_{10}(\orob/\re)}
\newcommand{\offset} {\mathcal{C}}
\newcommand{\scatter} {\sigma_\mathcal{R}}
\newcommand{\normal} {\mathcal{N}}

%% file: forecasting.tex
\subsection{Data Requirements}

The principal objective of this work is to compute self-consistent and
homogeneous a-posteriori distributions for the predicted mass of each KOI. The
predicted mass of each KOI will be solely determined by its radius and the
empirical, probabilistic forecasting model of \citet{chen:2017}. Because of
this conditional relationship, then mass and radius will certainly be
covariant, along with any other derived terms based on these quantities. We
therefore aim to derive the joint posteriors for all the parameters of
interest, which will encode any resulting covariances.

To accomplish this goal, we first require posterior distributions for the KOI
radii. The transit light curve of each KOI enables a measurement of the
planet-to-star radius ratio, $p$, with the precision depending upon photometric
quality, number and duration of observed transit events and modest degeneracies
with other covariant terms describing the transit shape \citep{carter:2008}.
With the quantity $p$ in-hand, it may be combined with the stellar radius,
$R_{\star}$, to infer the true planetary size, $R_P$. Therefore, to make
progress we need homogenous posterior probability distributions for 1) basic
transit parameters of each KOI 2) fundamental stellar properties of each KOI.

\subsection{Transit Posteriors}

Basic transit parameters of each KOI are provided in the NASA Exoplanet Archive
(\href{http://exoplanetarchive.ipac.caltech.edu}{NEA}, \citealt{NEA}), but
these are summary statistics rather than full posterior distributions, as
required for this work. Fortunately, \citet{rowe:2015} provide posterior
distributions for almost every KOI, with 100,000 samples for each obtained via
a Markov Chain Monte Carlo (MCMC) regression of a \citet{mandel:2002} light
curve model to the \textit{Kepler} photometric times series (we direct the
reader to \citealt{rowe:2015} for details). We downloaded all available
posteriors and found 7106 object files.

Except for two KOIs, all of these objects appear in the currently listed NEA
database (comprising of 9564 KOIs), with the exceptions being KOI-1168.02 and
KOI-1611.02. We deleted these two objects from our sample in what follows,
giving us 7104 KOIs. The other 2460 KOIs were not modeled by \citet{rowe:2015}
and thus are not considered further in what follows.

\subsection{Stellar Posteriors}

For stellar properties, we again note that summary statistics are available
on NEA but posteriors are not directly available. As one of our criteria is a
homogenous set of of posteriors, and we wish to calculate masses for as many
KOIs as possible, inferences for particular subsets of the KOI database
are not as useful for this work. Instead, we used the publicly available
posteriors from \citet{mathur:2017} who use information such as colors,
spectroscopy and asteroseismology to fit Dartmouth isochrone models
\citep{dotter:2008} for each \textit{Kepler} star, giving 40,000 posterior
samples per star.

We attempted to download stellar posteriors for all 7104 KOIs in the
\citet{rowe:2015} database, but found that posteriors were missing for 93
KOIs spanning 87 stars. These KOIs are flagged with a ``1'' in
Table~1 and were not considered further for analysis.

\input{flagtable.tex}

\subsection{KOI Radii Posteriors}

We next combined these distributions together to generate fair realizations of
the KOI radii. We do this by consecutively stepping through each row of the
\citet{mathur:2017} samples and drawing a random row from the corresponding
\citet{rowe:2015} posterior samples for $p$. This is possible because the
two posteriors are completely independently and share no covariances.

This process results in 40,000 fair realizations for the radius of each KOI,
where we report the radii in units of Earth radii ($R_{\oplus}=6378.1$\,km),
which are made available in the public posteriors available at \httplink (https://github.com/chenjj2/forecasts).

During this process, we found 38 KOIs could not locate the corresponding
MCMC file for the \citet{rowe:2015} transit parameters. It is unclear why
these were missing but given their relativity small number, we simply flag
them with a ``2'' in Table~1 and do not consider them further
for analysis. At this point, we are left with 6973 KOIs for which we have
been able to derive a radius posterior, of which 2283 are dispositioned as
``CONFIRMED'' on NEA, 1665 are ``CANDIDATE'' and 3025 are
``FALSE POSITIVE'' (these dispositions are also provided in 
Table~2).

\subsection{Predicting Masses with \forecaster}

The next step is to take the radius posterior of each KOI and, row-by-row,
predict a corresponding predicted mass with \forecaster. Whilst we direct
the reader to \citet{chen:2017} for a full description of \forecaster, we
here briefly describe the model and how it was calibrated.
We also direct the reader to \citet{wolfgang:2016} for their prior and
alternative discussion of probabilistic mass-radius relations.

\forecaster\ is fundamentally probabilistic, by which we mean that the model
includes intrinsic dispersion in the mass-radius relation to account for
additional variance beyond measurement uncertainties. This dispersion
represents the variance observed in nature itself. In contrast, a
deterministic model, in the case of a simple power-law, would be given by

\begin{equation}
\frac{\orob}{\re} =  \powerconst \Big(\frac{\omob}{\me}\Big)^\slope,
\label{power_deter}
\end{equation}

where $\orob$ \& $\omob$ are the mass and radius of the object and
$\powerconst$ \& $\slope$ are the parameters of the power-law. In this
relation, a single radius value corresponds to a single mass, once the shape
parameters have been trained. However, in reality of course, two planets could
have the same masses but different radii due to, for example, distinct
compositions. The power of the probabilistic formalism is to catch the 
intrinsic variance between different planets, essentially adding an extra
noise term which absorbs our ignorance of the true model.  Following
\citet{chen:2017}, we substitute $\logm$ and $\logr$ represent $\logmeq$ and
$\logreq$ respectively, to write our probabilistic model as

\begin{equation}
\logr \sim \normal (\mu = \offset+\logm \times \slope, \sigma = \scatter)
\label{linear_prob}
\end{equation}

where $\offset = \log_{10} \powerconst$ and $\normal(\mu,\sigma)$ is a normal
distribution. The model is also empirical because it is derived by fitting the
above mass-radius relation conditioned on a sample of 316 well-constrained
objects, with detailed tests in \citet{chen:2017} demonstrating that the sample
provides an unbiased training set.

Another characteristic of \forecaster\ is that it was only trained within a
specific (albeit broad) mass range, from $3\times10^{-4}\ M_\oplus$ to
$3\times10^{5}\ M_\oplus$, corresponding to dwarf planets to late-type stars.
However, some extreme KOIs have radii which fall outside of the expected
corresponding radius range. To enable a homogeneous analysis, we simply
extend the first and last part of the broken power-law relation, so that it
could cover a semi-infinite interval. In practice, this is only necessary for
very large KOIs exceeding a Solar radius, for which the KOI cannot be a planet
in any case, and thus these extrapolations simply highlight the unphysical
nature of these rare cases.

As discussed earlier, the model was applied to each radius posterior
sample for each KOI (278.92 million individual forecasts). It is
important to stress that the probabilistic nature of \forecaster\
means that re-running the same script again on the same KOI would
lead to a slightly different set of posterior samples for the planetary
mass, although they would still of course be fair and representative
samples.

\begin{figure}
\begin{center}
\includegraphics[width=8.4cm,angle=0,clip=true]{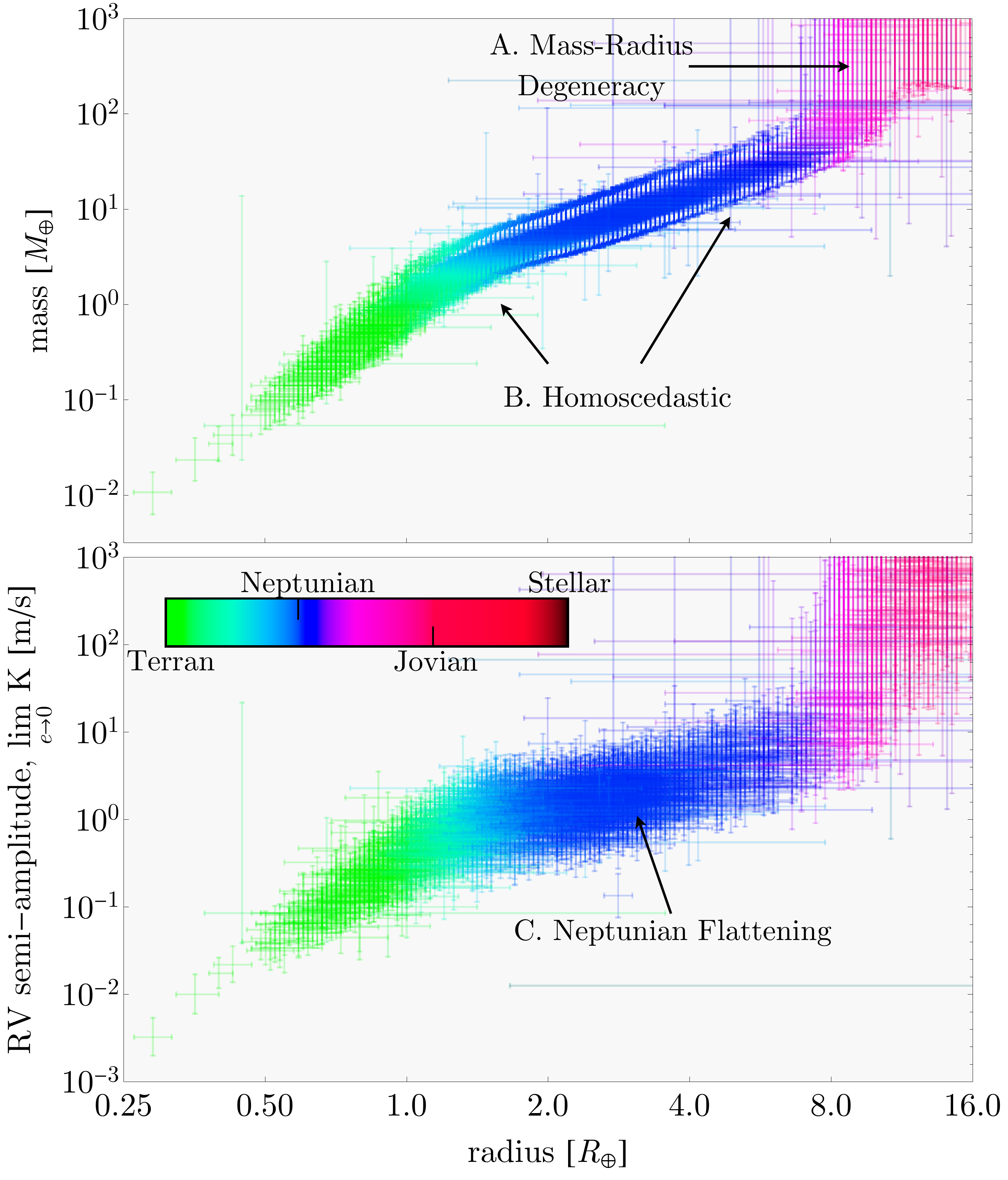}
\caption{
Forecasted masses (top) and radial velocity semi-amplitudes (bottom) for a
circular orbit, as a function of the observed KOI radius. We here only show
objects with a NEA disposition of being either a candidate or validated planet.
%Colors follows the CIE 1931 XYZ color space, where Y is the probability that
%the KOI is Terran, Z is that of being Neptunian and X is that of being Jovian.
Error bars depict the 68.3\% credible intervals.
}
\label{fig:RMK}
\end{center}
\end{figure}

The 68.3\% credible intervals of the forecasted masses are depicted in
upper panel of Figure~\ref{fig:RMK} and listed along with the 95.5\%
credible intervals in Table~2.

To perform the analysis, we used the function "Rpost2M" in \forecaster. To be clearer,
a summary of the steps taken in the function is listed as below.
1. generates a grid of mass in the allowed mass range
2. calculates the probabilities of the measured radius given each mass in the grid
3. use the above probabilities as weights to redraw mass
For more details, we direct the reader to the original \forecaster paper \citet{chen:2017}.

\input{finaltable.tex}

%% file: flagtable.tex
\begin{table}
\label{tab:flags}
\caption{
Data flags assigned to 7104 KOIs considered in this analysis.
``0'' denotes no problems, ``1'' denotes that stellar posteriors
were missing and ``2'' denotes transit posteriors were missing.
Only a portion of the table is shown here, the full version is
available in the online version.
}
\centering % used for centering table
\begin{tabular}{cc} % centered columns (7 columns)
\hline
KOI & data flag \\ [0.5ex] % inserts table
%heading
\hline
0001.01	&	0 \\
0002.01	&	0 \\
0003.01	&	0 \\
0004.01	&	0 \\
0005.01	&	0 \\
0005.02	&	2 \\
0006.01	&	0 \\
0007.01	&	0 \\
0008.01	&	0 \\
0009.01	&	0 \\
0010.01	&	0 \\
\vdots & \vdots \\ [1ex]
\hline %inserts single line
\end{tabular}
\end{table}

%% file: finaltable.tex
\begin{landscape}
\begin{table}
\label{tab:final}
\caption{
Final predicted properties for 6973 KOIs using \forecaster. Quoted values are the [$-2,-1,0,+1,+2$]\,$\sigma$
credible intervals. First letter of the flag column gives modal planet classification (``T''=Terran, ``N''=Neptunian,
``J''=Jovian \& ``S''=Stellar). Second letter of the flag column denotes NEA KOI disposition where ``f'' is a false-positive, ``c'' is a candidate and ``v'' is validated/confirmed. Only a portion of the table is shown here, the full version is available in the online version.
}
%\centering % used for centering table
\begin{tabular}{lllllll} % centered columns (7 columns)
\hline
KOI & flag & $\log_{10}(R_P$\,$[R_{\oplus}])$ & $\log_{10}(M_P$\,$[M_{\oplus}])$ & $\log_{10}(K\,[\mathrm{m}\,\mathrm{s}^{-1}])$ & $\log_{10}(\rho_P\,[\mathrm{g}\,\mathrm{cm}^{-3}])$ & $\log_{10}(g_P\,[\mathrm{cm}\,\mathrm{s}^{-2}])$ \\ [0.5ex] % inserts table
%heading
\hline
0001.01	&	Jv	&	[1.08,1.10,1.11,1.13,1.15]	&	[1.83,2.33,3.20,4.14,4.53]	&	[1.51,2.02,2.88,3.82,4.19]	&	[-0.77,-0.27,0.60,1.55,1.93]	&	[2.59,3.09,3.96,4.91,5.29]	\\
0002.01	&	Jv	&	[1.16,1.18,1.20,1.22,1.24]	&	[1.86,2.23,2.95,4.51,4.67]	&	[1.44,1.82,2.54,4.08,4.22]	&	[-1.01,-0.63,0.09,1.63,1.79]	&	[2.44,2.82,3.54,5.09,5.24]	\\
0003.01	&	Nv	&	[0.66,0.67,0.69,0.70,0.71]	&	[0.81,1.07,1.32,1.58,1.83]	&	[0.45,0.71,0.96,1.21,1.46]	&	[-0.49,-0.23,0.01,0.26,0.51]	&	[2.43,2.69,2.94,3.19,3.44]	\\
0004.01	&	Jc	&	[0.85,0.92,1.00,1.08,1.17]	&	[1.34,1.72,3.04,4.13,4.46]	&	[0.90,1.27,2.55,3.64,3.95]	&	[-0.76,-0.39,0.64,1.83,2.22]	&	[2.47,2.82,3.94,5.10,5.43]	\\
0005.01	&	Nc	&	[0.83,0.85,0.87,0.89,0.91]	&	[1.13,1.40,1.67,2.00,4.28]	&	[0.64,0.91,1.18,1.51,3.77]	&	[-0.73,-0.46,-0.19,0.14,2.37]	&	[2.38,2.65,2.92,3.25,5.51]	\\
0006.01	&	Sf	&	[1.21,1.38,1.60,1.91,2.10]	&	[2.10,4.77,5.05,5.39,5.48]	&	[1.83,4.46,4.69,4.96,5.03]	&	[-1.05,0.23,0.89,1.30,1.56]	&	[2.48,4.43,4.79,4.99,5.12]	\\
0007.01	&	Nv	&	[0.58,0.60,0.62,0.65,0.67]	&	[0.70,0.96,1.22,1.47,1.73]	&	[0.31,0.57,0.82,1.07,1.32]	&	[-0.42,-0.16,0.08,0.34,0.59]	&	[2.45,2.71,2.96,3.21,3.46]	\\
0008.01	&	Nf	&	[0.19,0.22,0.26,0.32,0.38]	&	[0.26,0.44,0.66,0.90,1.17]	&	[0.06,0.24,0.45,0.70,0.97]	&	[0.15,0.37,0.60,0.84,1.09]	&	[2.71,2.90,3.11,3.35,3.60]	\\
0009.01	&	Jf	&	[0.99,1.06,1.15,1.27,1.37]	&	[1.77,2.26,3.27,4.56,4.80]	&	[1.32,1.80,2.83,4.08,4.24]	&	[-1.07,-0.48,0.62,1.62,1.97]	&	[2.41,2.92,4.00,5.10,5.30]	\\
0010.01	&	Jv	&	[1.11,1.14,1.17,1.20,1.23]	&	[1.85,2.27,3.03,4.23,4.63]	&	[1.43,1.85,2.61,3.81,4.17]	&	[-0.93,-0.50,0.27,1.50,1.85]	&	[2.49,2.92,3.69,4.91,5.26]	\\
0011.01	&	Sf	&	[0.77,0.93,1.18,1.51,1.76]	&	[1.32,1.86,3.78,4.95,5.24]	&	[0.97,1.51,3.43,4.52,4.75]	&	[-1.04,-0.35,0.85,1.53,2.08]	&	[2.41,2.87,4.62,5.05,5.33]	\\
0012.01	&	Jv	&	[1.09,1.13,1.17,1.21,1.27]	&	[1.86,2.27,3.04,4.31,4.67]	&	[1.14,1.56,2.34,3.59,3.91]	&	[-0.96,-0.51,0.29,1.55,1.85]	&	[2.47,2.91,3.70,4.99,5.26]	\\
0013.01	&	Jv	&	[1.05,1.12,1.21,1.32,1.50]	&	[1.87,2.29,3.35,4.71,4.95]	&	[1.42,1.85,2.93,4.21,4.36]	&	[-1.13,-0.55,0.61,1.57,1.85]	&	[2.38,2.88,4.02,5.10,5.25]	\\
0014.01	&	Nf	&	[0.55,0.62,0.69,0.80,0.90]	&	[0.77,1.05,1.35,1.65,2.04]	&	[0.29,0.57,0.85,1.14,1.52]	&	[-0.57,-0.28,-0.00,0.27,0.65]	&	[2.42,2.69,2.94,3.20,3.56]	\\
0015.01	&	Sf	&	[1.45,1.70,2.02,2.28,2.46]	&	[2.01,4.96,5.40,5.48,5.48]	&	[1.42,4.40,4.75,4.83,4.86]	&	[-4.55,-0.66,0.12,0.79,1.25]	&	[0.14,3.87,4.39,4.75,4.97]	\\
0016.01	&	Sf	&	[0.79,1.01,1.64,1.94,2.13]	&	[1.36,2.49,5.09,5.42,5.48]	&	[1.06,2.15,4.70,4.97,5.02]	&	[-0.76,-0.09,0.59,1.22,1.87]	&	[2.53,3.29,4.63,4.92,5.19]	\\
0017.01	&	Jv	&	[1.08,1.10,1.13,1.15,1.19]	&	[1.82,2.32,3.14,4.14,4.56]	&	[1.44,1.93,2.76,3.75,4.14]	&	[-0.82,-0.33,0.51,1.52,1.91]	&	[2.56,3.04,3.88,4.89,5.28]	\\
0018.01	&	Jv	&	[1.12,1.14,1.18,1.21,1.24]	&	[1.86,2.27,3.01,4.29,4.64]	&	[1.41,1.81,2.56,3.83,4.15]	&	[-0.95,-0.52,0.23,1.53,1.84]	&	[2.49,2.90,3.65,4.95,5.25]	\\
0019.01	&	Sf	&	[0.96,1.12,1.32,1.45,1.56]	&	[1.77,2.39,4.67,4.88,5.03]	&	[1.55,2.16,4.40,4.59,4.70]	&	[-1.30,-0.40,1.23,1.52,1.88]	&	[2.28,3.00,4.92,5.08,5.23]	\\
0020.01	&	Jv	&	[1.11,1.16,1.23,1.30,1.38]	&	[1.86,2.25,3.25,4.68,4.82]	&	[1.43,1.82,2.82,4.20,4.31]	&	[-1.18,-0.65,0.42,1.59,1.79]	&	[2.35,2.81,3.86,5.11,5.23]	\\
0021.01	&	Sf	&	[1.22,1.35,1.51,1.69,1.87]	&	[2.04,4.73,4.95,5.16,5.37]	&	[1.64,4.28,4.46,4.63,4.78]	&	[-1.22,0.63,1.07,1.37,1.58]	&	[2.35,4.66,4.88,5.02,5.14]	\\
0022.01	&	Jv	&	[1.01,1.05,1.09,1.16,1.23]	&	[1.74,2.29,3.24,4.20,4.57]	&	[1.22,1.76,2.71,3.65,4.00]	&	[-0.84,-0.31,0.70,1.67,2.00]	&	[2.53,3.05,4.04,5.02,5.32]	\\
0023.01	&	Jf	&	[1.14,1.19,1.24,1.30,1.38]	&	[1.86,2.23,3.24,4.69,4.82]	&	[1.37,1.74,2.76,4.16,4.26]	&	[-1.19,-0.71,0.35,1.60,1.77]	&	[2.34,2.76,3.81,5.12,5.22]	\\
0024.01	&	Jf	&	[0.83,0.89,0.98,1.09,1.20]	&	[1.28,1.64,2.74,4.10,4.47]	&	[1.03,1.38,2.46,3.82,4.17]	&	[-0.78,-0.42,0.29,1.80,2.24]	&	[2.44,2.77,3.60,5.08,5.44]	\\
0025.01	&	Sf	&	[1.22,1.29,1.37,1.49,1.62]	&	[1.93,2.74,4.78,4.93,5.09]	&	[1.55,2.38,4.35,4.46,4.57]	&	[-1.37,-0.32,1.29,1.50,1.64]	&	[2.23,3.19,4.97,5.09,5.17]	\\
0026.01	&	Jf	&	[0.94,1.00,1.09,1.23,1.41]	&	[1.61,2.18,3.33,4.41,4.84]	&	[1.01,1.53,2.70,3.77,4.12]	&	[-0.97,-0.39,0.81,1.68,2.08]	&	[2.45,2.95,4.15,5.09,5.35]	\\
0027.01	&	Sf	&	[1.83,1.89,1.97,2.09,2.23]	&	[5.27,5.36,5.44,5.48,5.48]	&	[4.95,5.01,5.05,5.08,5.10]	&	[-0.46,-0.04,0.27,0.45,0.58]	&	[4.01,4.28,4.48,4.59,4.67]	\\
0028.01	&	Sf	&	[1.68,1.73,1.82,1.94,2.06]	&	[5.10,5.19,5.30,5.43,5.48]	&	[4.62,4.68,4.76,4.83,4.87]	&	[0.01,0.33,0.58,0.74,0.87]	&	[4.32,4.52,4.64,4.74,4.81]	\\
0031.01	&	Sf	&	[1.59,1.65,1.72,1.80,1.88]	&	[5.00,5.09,5.19,5.30,5.39]	&	[4.72,4.79,4.85,4.92,4.98]	&	[0.45,0.60,0.76,0.91,1.04]	&	[4.56,4.65,4.73,4.81,4.89]	\\
0033.01	&	Sf	&	[1.58,1.62,1.68,1.74,1.78]	&	[4.98,5.05,5.13,5.22,5.30]	&	[4.84,4.89,4.94,4.98,5.03]	&	[0.62,0.72,0.84,0.96,1.06]	&	[4.63,4.70,4.77,4.84,4.91]	\\
0041.01	&	Nv	&	[0.30,0.32,0.35,0.37,0.40]	&	[0.33,0.53,0.77,1.01,1.26]	&	[-0.24,-0.04,0.18,0.42,0.67]	&	[0.02,0.24,0.47,0.71,0.96]	&	[2.63,2.84,3.06,3.30,3.55]	\\
0041.02	&	Nv	&	[0.06,0.09,0.11,0.15,0.19]	&	[-0.00,0.17,0.35,0.59,0.86]	&	[-0.49,-0.31,-0.14,0.10,0.36]	&	[0.41,0.58,0.74,0.97,1.23]	&	[2.78,2.95,3.10,3.34,3.60]	\\
0041.03	&	Nv	&	[0.12,0.14,0.17,0.20,0.25]	&	[0.14,0.31,0.49,0.73,0.98]	&	[-0.59,-0.42,-0.24,-0.00,0.25]	&	[0.36,0.54,0.72,0.94,1.20]	&	[2.80,2.96,3.14,3.36,3.62]	\\
0042.01	&	Nv	&	[0.35,0.37,0.39,0.41,0.43]	&	[0.37,0.59,0.83,1.07,1.33]	&	[-0.28,-0.06,0.16,0.41,0.67]	&	[-0.04,0.17,0.41,0.65,0.91]	&	[2.59,2.81,3.04,3.29,3.55]	\\
0044.01	&	Jf	&	[0.98,1.03,1.10,1.19,1.31]	&	[1.71,2.26,3.27,4.29,4.71]	&	[0.91,1.43,2.45,3.47,3.81]	&	[-0.92,-0.37,0.70,1.67,2.03]	&	[2.48,2.99,4.06,5.08,5.33]	\\
0046.01	&	Nv	&	[0.58,0.64,0.70,0.78,0.85]	&	[0.80,1.08,1.36,1.64,1.93]	&	[0.42,0.70,0.97,1.24,1.52]	&	[-0.55,-0.28,-0.01,0.24,0.53]	&	[2.43,2.68,2.94,3.19,3.45]	\\
0046.02	&	Tv	&	[-0.11,-0.05,0.01,0.09,0.17]	&	[-0.52,-0.24,0.06,0.40,0.73]	&	[-0.98,-0.70,-0.40,-0.08,0.23]	&	[0.40,0.57,0.73,0.94,1.31]	&	[2.62,2.81,3.01,3.24,3.58]	\\
0048.01	&	Nf	&	[2.07,2.27,2.53,2.86,3.11]	&	[1.91,2.04,2.13,5.48,5.48]	&	[1.25,1.38,1.48,4.64,4.67]	&	[-6.48,-5.73,-4.74,-0.59,-0.01]	&	[-1.12,-0.62,0.03,3.92,4.31]	\\
0049.01	&	Nv	&	[0.41,0.43,0.47,0.51,0.55]	&	[0.47,0.71,0.95,1.22,1.48]	&	[-0.03,0.19,0.44,0.70,0.96]	&	[-0.19,0.04,0.29,0.54,0.80]	&	[2.53,2.76,3.01,3.26,3.51]	\\
0052.01	&	Jf	&	[1.18,1.22,1.26,1.31,1.41]	&	[1.86,2.22,3.40,4.71,4.86]	&	[1.56,1.91,3.10,4.35,4.46]	&	[-1.22,-0.79,0.44,1.60,1.73]	&	[2.31,2.71,3.93,5.13,5.21]	\\
0061.01	&	Sf	&	[1.11,1.41,1.75,2.01,2.18]	&	[2.14,4.81,5.21,5.46,5.48]	&	[1.85,4.49,4.82,4.99,5.03]	&	[-0.86,0.04,0.61,1.19,1.56]	&	[2.64,4.31,4.65,4.93,5.11]	\\
0063.01	&	Nv	&	[0.71,0.73,0.77,0.81,0.85]	&	[0.95,1.21,1.46,1.72,1.99]	&	[0.45,0.71,0.97,1.22,1.49]	&	[-0.61,-0.35,-0.09,0.15,0.42]	&	[2.41,2.67,2.92,3.17,3.43]	\\
0064.01	&	Nc	&	[0.81,0.86,0.91,0.96,1.03]	&	[1.18,1.48,1.82,3.89,4.37]	&	[0.86,1.15,1.48,3.54,4.00]	&	[-0.74,-0.45,-0.11,1.79,2.32]	&	[2.40,2.69,3.02,4.99,5.49]	\\
0066.01	&	Sf	&	[2.11,2.19,2.28,2.35,2.42]	&	[1.94,2.10,5.47,5.48,5.48]	&	[1.69,1.85,5.11,5.13,5.14]	&	[-4.43,-4.15,-0.63,-0.36,-0.11]	&	[0.21,0.42,3.90,4.07,4.24]	\\
\vdots & \vdots & \vdots & \vdots & \vdots & \vdots & \vdots \\ [1ex]
\hline %inserts single line
\end{tabular}
\end{table}
\end{landscape}

%% file: implications.tex
\subsection{Densities, Gravities and RV Amplitudes}

With a joint posterior for the both the stellar and planetary fundamental
parameters in hand, we can compute other derived quantities of interest,
such as surface gravity, bulk density and radial velocity (RV) amplitude
(these are given in Table~2). Once again, we stress that these
values should be treated as forecasts and not measurements.

We have computed forecasted masses for all KOIs where available, irrespective
of whether the KOI is a known false-positive or confirmed planet.
False-positives frequently have extreme radii associated with them, giving
rise to anomalous derived parameters. For example, in the case of KOI-5385.01,
a known false-positive, we obtain a radius of
$R_P=1010_{-500}^{+450}$\,$R_{\odot}$ (for comparison NEA report
$R_P=62_{-5}^{+20}$\,$R_{\odot}$), implying the \citet{rowe:2015}
MCMC chains ultimately diverged to a very high $p$. We caution that this
effect also appears for some KOIs which are not dispositioned as
false-positives, for example in the case of KOI-3891.01, which is listed as
a candidate on NEA, we obtain $R_P = 126_{-89}^{+132}$\,$R_{\odot}$
(NEA also report an anomalously large radius of $R_P = 
0.81_{-0.52}^{+0.28}$\,$R_{\odot}$). Despite this, these cases are
straight-forward to identify and essentially represent cases where the MCMC
diverged, indicative of poor quality light curves or false-positives.

The density forecasts are computed assuming a spherical shape for the
planet and are listed in Table~2. Density forecasts are
particularly useful in cases where one wishes to predict the Roche radius
around KOIs for estimating potential ring radii \citep{zuluaga:2015}.

The surface gravity forecasts are also made assuming spherical planets.
These estimates are particularly important for climate and atmospheric
modeling \citep{heng:2011}, as well as predicting the scale height of
exoplanetary atmospheres \citep{seager:2000} when planning observational
campaigns.

Finally, the radial velocity amplitudes are computed assuming
$i\simeq\pi/2$, as appropriate for transiting planets,and additionally
under the explicit assumption of a circular orbit, such that

\begin{align}
\lim_{e\to0} K &= \Bigg( \frac{2 \pi G}{P} \Bigg)^{1/3} \frac{M_P}{(M_{\star}+M_P)^{2/3}}.
\end{align}

We do not assume $M_P \ll M_{\star}$ and ensure that the uncertainties in the
stellar mass are propagated correctly into the calculation of $K$, although
we assume negligible uncertainty in orbital period, $P$. Circularity
is assumed to provide a tight fiducial value rather than marginalizing
over poorly constrained eccentricity distributions for these worlds. These
predictions should help observers plan which KOIs have potentially
detectable signatures from the ground.

We highlight that our predicted masses are calculated homogeneously for
all KOIs, regardless as to whether the system is known to exhibit strong
transit timing variations (TTV) or not. This is relevant since planetary masses
detected through the TTV method, rather than the RV method, are subject to
distinct selection effects leading to the possibility of ostensibly distinct
mass distributions \cite{steffen:2016}. Further, the original \citet{chen:2017}
calibration of \forecaster\ is dominated by planets with masses measured through
RVs. As this issue continues to be investigated, it will be interesting to test
then if subsequent TTV detected masses are offset from the predictions in this
work.

\subsection{Observed Patterns}
\label{sub:patterns}

From studying the results, we highlight three noticeable patterns, which are
highlighted in Figure~\ref{fig:RMK}. Observation A: the masses (and indeed RV
predictions) follow a relatively tight relation up to $\simeq10$\,$R_{\oplus}$
after which point the uncertainties greatly increase. Naively, one might assume
that the uncertainties should be smaller here, since larger planets give rise
to deeper transits and thus we acquire a higher relative signal-to-noise.
However, at around a Jupiter-radius, degeneracy pressure leads to an almost flat
mass-radius relation all the way until deuterium burning starts in the stars,
which occurs at $M=0.008_{-0.0072}^{+0.0081}$\,$M_{\odot}$ \citep{chen:2017}.
As a result of this degeneracy, mass predictions end up spanning almost the
entire Jovian range, leading to much larger credible intervals.

Observation B: we note that the radius-mass diagram in Figure~\ref{fig:RMK}
shows that the credible interval of the Neptunians and even many of the Terrans
have approximately the same width (in logarithmic space) i.e. the logarithmic
variance is homoscedastic. Ultimately, these uncertainties are a combination of
the measurement error in the observed radii and the intrinsic dispersion in the
mass-radius relation. Accordingly, the fact that the uncertainties appear
approximately homoscedastic indicates that they are dominated by the intrinsic
dispersion term, rather than the measurement errors on $R$. This,
in turn, implies that it will not be possible to forecast noticeably
reduced credible intervals in the future by using more precise planetary
radii (for example by using more precise stellar radii e.g.
\citealt{CKS:2017}). Therefore, the only way to \textit{forecast} improved
credible intervals would be to train a revised \forecaster\ model,
perhaps using additional variables treated as latent in the current
version (e.g. insolation). Until that happens though, the forecasted
masses for $R\lesssim10$\,$R_{\oplus}$ presented here can be treated
as the most precise possible regardless of observational error.

\begin{figure}
\begin{center}
\includegraphics[width=8.4cm,angle=0,clip=true]{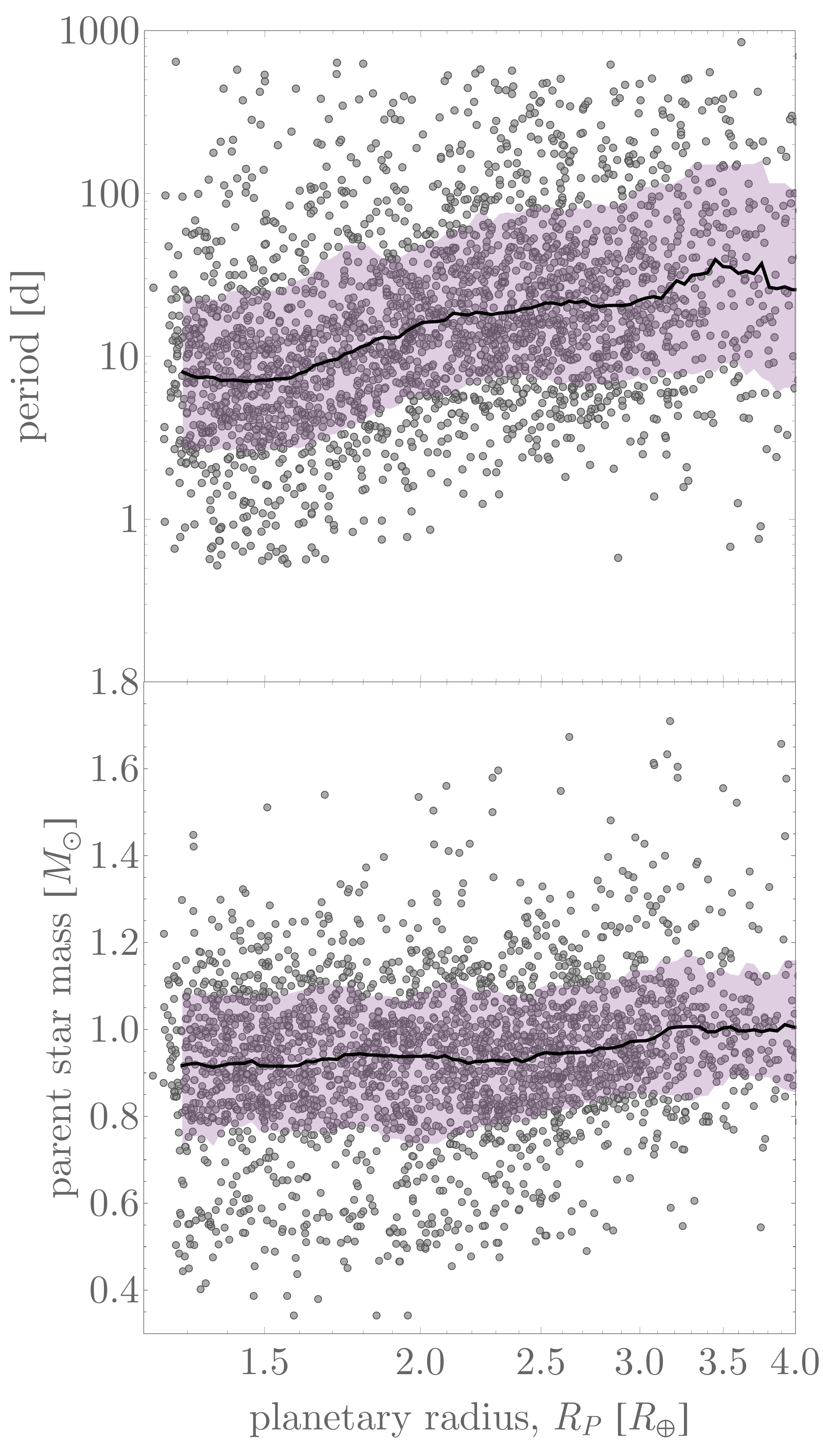}
\caption{Orbital periods (top) and host star masses (bottom) as
a function of planetary radius for the 2936 KOIs with modal
class probability of being Neptunian and not dispositioned as
a false-positive on NEA. The trend between period and radius
is most easily explained as being due to detection bias, which
in turn gives rise to the RV plateau seen for the Neptunians in
Figure~\ref{fig:RMK}. Median (black line) and 68.3\% credible
interval (purple) shown using a top-hat smoothing kernel of
0.14\,dex.
}
\label{fig:correl}
\end{center}
\end{figure}

Observation C: we observe a flattening in the gradient of the predicted RV
amplitude as a function of planetary radius for the Neptunian planets
(see Figure~\ref{fig:RMK}). This is somewhat surprising because the RV
amplitude is proportional to planetary mass, yet the radius-mass diagram
shows a steeper dependency in the Neptunian range. The implication from
this is that large Neptunes detected by \textit{Kepler} are predicted to
have almost the same RV amplitude as small, detected Neptunes. Given that
$M_P \ll M_{\star}$ in this regime, and we've assumed circular orbits,
there are only two ways to explain this: i) the larger Neptunes tend to
have longer orbital periods ii) the larger Neptunes tend to have higher
mass host stars (or a combination of the two effects). Both effects are
also detection biases for \textit{Kepler} \citep{sandford:2016}, since
long-period planets transit less frequently and high-mass stars tend to be
larger, giving rise to smaller transit depths. To investigate which effect
dominates, we plot the orbital periods and host star masses as a function
of planetary radii for the Neptunian planets in Figure~\ref{fig:correl}.

Figure~\ref{fig:correl} reveals that there is indeed an apparent trend
between planetary radius and orbital period for the \textit{Kepler}
Neptunes, which is most easily explained as being due to detection bias.
We find no such trend for stellar mass, although \textit{Kepler} had a
strong selection bias towards Solar-like stars and thus the sample is
limited for low-mass stars.

\subsection{Promising Small Planets for RV follow-up}
\label{sub:opportunities}

We here demonstrate perhaps the most useful application of this work,
for identifying promising small planets for RV follow-up. In this work,
we have applied \forecaster\ to \textit{Kepler} planets but this
should be treated as a demonstration of what could be done with the
upcoming \textit{TESS} survey too \citep{ricker}. In particular,
the narrow-but-deep nature of \textit{Kepler} combined with the
fact RV amplitudes are enhanced for low-mass star means that the most
favorable targets (in terms of absolute RV amplitude) will be around
faint stars. The all-sky nature of \textit{TESS} will lead to brighter
targets and here \forecaster\ can clearly play a direct role in
identifying promising follow-up targets.

\input{bestones.tex}

\begin{figure}
\begin{center}
\includegraphics[width=8.4cm,angle=0,clip=true]{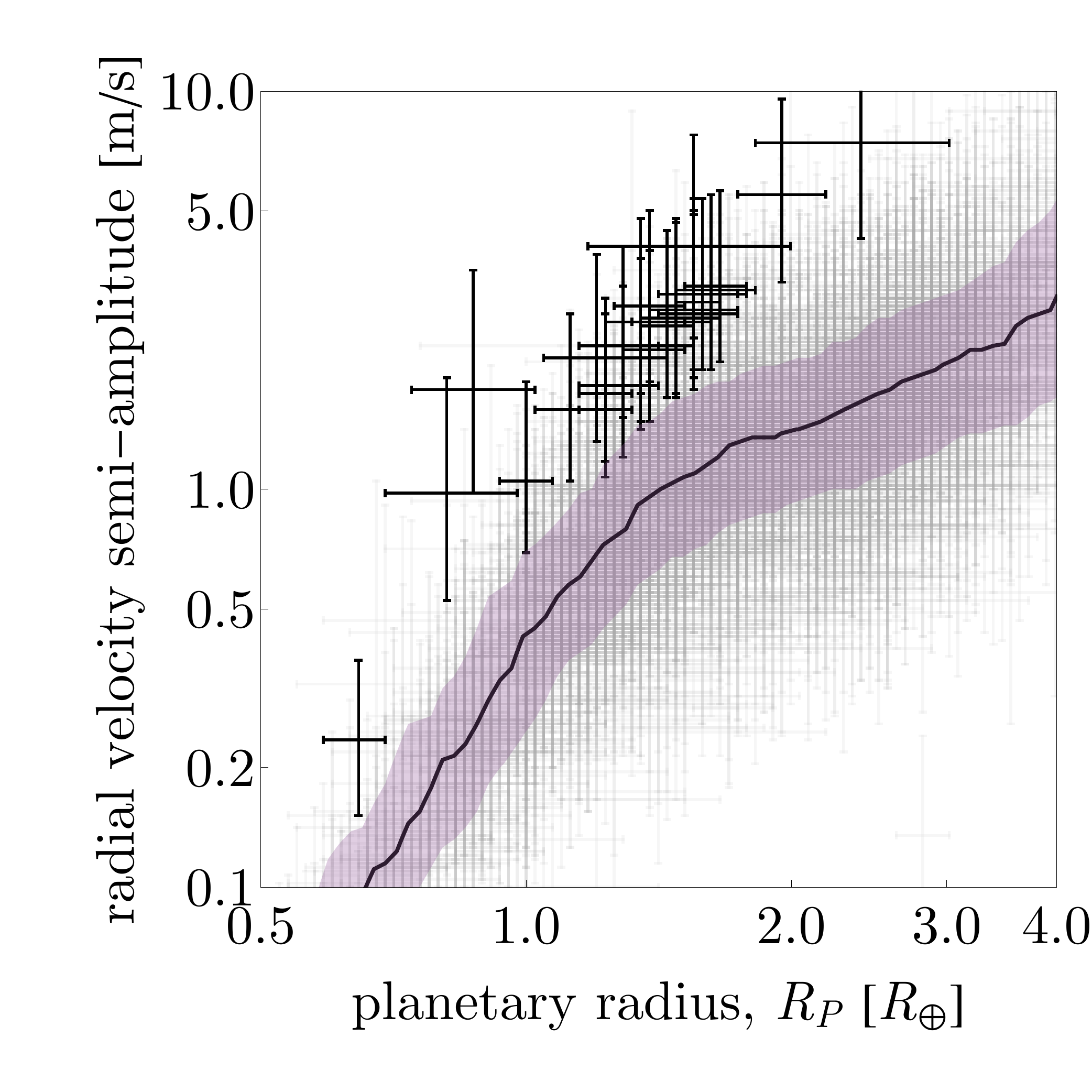}
\caption{
Forecasted radial velocity amplitudes for small planets. The
colored region contains the 68.3\% credible interval of maximum
likeihood forecasted $K$ values from a moving window. Points
plotted use 68.3\% uncertainties and the black points are those
more than 1.64\,$\sigma$ (90\%) above the moving median (black).
}
\label{fig:bestones}
\end{center}
\end{figure}

We list 28 promising targets in Table~3, where we haved
down-selected on KOIs not dispositioned as false-positives on NEA, with
radii between one half to four times that of the Earth and with unusually high
RV amplitude predictions. For the latter criteria, we specifically
used $\hat{K}-K_{\mathrm{med}}>1.64 (\hat{K}-K_{\mathrm{lower}})$,
where $\hat{K}$ is the median forecast for $K$, $K_{\mathrm{lower}}$
is the 15.9\% lower quantile of the forecasted $K$ distribution and
$K_{\mathrm{med}}$ is the running median of the median forecasts
of $K$ with the same window size as used before (note that
$1.64$\,$\sigma=90$\%). These KOIs are visualized versus the
ensemble of small KOIs in Figure~\ref{fig:bestones}.

As can be seen from Table~3, this sample of 28 KOIs
all have \textit{Kepler}-bandpass magnitudes fainter than $K_P>14$.
The 1.35\,$R_{\oplus}$ planetary candidate KOI-2119.01 is brightest
at $K_P=14.1$, for which we predict a $\sim2$\,m/s amplitude.
This sample of targets also highlights the great potential of
near-infrared spectrographs (e.g. \citealt{cursullo:2017}), where the
apparent magnitude is significantly brighter.

Despite their faintness, the very short-period nature of these
objects (as evident from Table~3) enables repeat monitoring
and potentially easier disentanglement of spurious signals originating from
the star \citep{pepe:2013,howard:2013}. Thus, despite the challenges,
these targets may be worth pursueing for \textit{Kepler} and indeed one
can expect this approach to yield far more suitable targets in the
\textit{TESS}-era.

%% file: bestones.tex
\begin{table}
\label{tab:bestones}
\caption{
List of planetary candidates with maximum likelihood radii between 0.5 to 4.0
Earth radii and unusually high forecasted RV semi-amplitudes.% These objects
%$\sim$m/s amplitudes for planets saddling the Terran-to-Neptunian transition
%and thus would be of great interest for follow-up, although the deep-sky nature
%of \textit{Kepler} make these faint in visible.
}
\centering % used for centering table
\begin{tabular}{cccccc} % centered columns (7 columns)
\hline
KOI & $R_P$\,$[R_{\oplus}]$ & $K$\,$[\mathrm{m/s}]$ & $M_{\star}$\,$[M_{\odot}]$ & $P_P$\,$[\mathrm{days}]$ & $K_P$ \\ [0.5ex] % inserts table
%heading
\hline
0596.01	&	$1.29_{-0.14}^{+0.12}$	&	$1.82_{-0.62}^{+1.42}$	&	$0.485_{-0.045}^{+0.039}$	&	$1.6827$ & 14.818 \\
0739.01	&	$1.45_{-0.10}^{+0.10}$	&	$2.57_{-0.87}^{+1.90}$	&	$0.533_{-0.037}^{+0.033}$	&	$1.2871$ & 15.488 \\
0936.02	&	$1.29_{-0.14}^{+0.12}$	&	$2.29_{-0.78}^{+1.78}$	&	$0.479_{-0.049}^{+0.039}$	&	$0.8930$ & 15.073 \\
0952.05	&	$1.38_{-0.12}^{+0.13}$	&	$2.88_{-1.02}^{+2.13}$	&	$0.506_{-0.041}^{+0.037}$	&	$0.7430$ & 15.801 \\
0961.01	&	$0.81_{-0.12}^{+0.16}$	&	$0.98_{-0.45}^{+0.93}$	&	$0.144_{-0.022}^{+0.030}$	&	$1.2138$ & 15.920 \\
0961.02	&	$0.87_{-0.13}^{+0.15}$	&	$1.78_{-0.80}^{+1.77}$	&	$0.144_{-0.022}^{+0.030}$	&	$0.4533$ & 15.920 \\
1202.01	&	$1.23_{-0.08}^{+0.09}$	&	$1.74_{-0.56}^{+1.28}$	&	$0.614_{-0.029}^{+0.030}$	&	$0.9283$ & 15.854 \\
1300.01	&	$1.12_{-0.10}^{+0.08}$	&	$1.58_{-0.54}^{+1.17}$	&	$0.540_{-0.034}^{+0.034}$	&	$0.6313$ & 14.285 \\
1367.01	&	$1.55_{-0.07}^{+0.11}$	&	$2.95_{-1.05}^{+2.06}$	&	$0.840_{-0.039}^{+0.037}$	&	$0.5679$ & 15.055 \\
1880.01	&	$1.35_{-0.12}^{+0.20}$	&	$2.29_{-0.81}^{+1.78}$	&	$0.553_{-0.031}^{+0.030}$	&	$1.1512$ & 14.440 \\
2119.01	&	$1.35_{-0.06}^{+0.10}$	&	$2.14_{-0.73}^{+1.66}$	&	$0.849_{-0.038}^{+0.039}$	&	$0.5710$ & 14.098 \\
2250.02	&	$1.66_{-0.15}^{+0.12}$	&	$3.24_{-1.15}^{+2.39}$	&	$0.812_{-0.059}^{+0.055}$	&	$0.6263$ & 15.622 \\
2347.01	&	$1.00_{-0.07}^{+0.07}$	&	$1.05_{-0.36}^{+0.81}$	&	$0.580_{-0.030}^{+0.029}$	&	$0.5880$ & 14.934 \\
2393.02	&	$1.23_{-0.08}^{+0.09}$	&	$1.58_{-0.51}^{+1.17}$	&	$0.790_{-0.036}^{+0.031}$	&	$0.7667$ & 14.903 \\
2409.01	&	$1.55_{-0.14}^{+0.19}$	&	$3.09_{-1.10}^{+2.28}$	&	$0.764_{-0.065}^{+0.080}$	&	$0.5774$ & 14.859 \\
2480.01	&	$1.35_{-0.12}^{+0.16}$	&	$2.63_{-0.89}^{+2.16}$	&	$0.558_{-0.048}^{+0.050}$	&	$0.6668$ & 15.745 \\
2493.01	&	$1.55_{-0.14}^{+0.19}$	&	$2.75_{-0.98}^{+2.14}$	&	$0.844_{-0.068}^{+0.066}$	&	$0.6631$ & 15.304 \\
2699.01	&	$1.62_{-0.14}^{+0.20}$	&	$3.16_{-1.17}^{+2.33}$	&	$0.853_{-0.077}^{+0.065}$	&	$0.5689$ & 15.230 \\
2704.01	&	$2.40_{-0.58}^{+0.62}$	&	$7.41_{-3.15}^{+6.08}$	&	$0.189_{-0.050}^{+0.068}$	&	$4.8712$ & 17.475 \\
2704.02	&	$1.55_{-0.37}^{+0.45}$	&	$4.07_{-1.67}^{+3.69}$	&	$0.189_{-0.050}^{+0.068}$	&	$2.9842$ & 17.475 \\
2735.01	&	$1.48_{-0.13}^{+0.18}$	&	$2.69_{-0.95}^{+2.09}$	&	$0.864_{-0.063}^{+0.054}$	&	$0.5588$ & 15.600 \\
2783.01	&	$0.65_{-0.06}^{+0.05}$	&	$0.23_{-0.08}^{+0.14}$	&	$0.518_{-0.032}^{+0.030}$	&	$0.5269$ & 14.694 \\
2793.02	&	$1.48_{-0.16}^{+0.14}$	&	$2.63_{-0.93}^{+2.05}$	&	$0.456_{-0.046}^{+0.043}$	&	$1.7668$ & 16.283 \\
2817.01	&	$1.55_{-0.17}^{+0.19}$	&	$2.82_{-1.04}^{+2.19}$	&	$0.805_{-0.074}^{+0.076}$	&	$0.6340$ & 15.760 \\
2842.01	&	$1.95_{-0.21}^{+0.24}$	&	$5.50_{-2.18}^{+4.05}$	&	$0.343_{-0.047}^{+0.044}$	&	$1.5654$ & 16.257 \\
2842.03	&	$1.58_{-0.17}^{+0.19}$	&	$3.09_{-1.10}^{+2.28}$	&	$0.343_{-0.047}^{+0.044}$	&	$3.0362$ & 16.257 \\
3119.01	&	$1.20_{-0.16}^{+0.18}$	&	$2.14_{-0.82}^{+1.75}$	&	$0.249_{-0.036}^{+0.045}$	&	$2.1844$ & 16.946 \\
4002.01	&	$1.38_{-0.09}^{+0.13}$	&	$2.24_{-0.76}^{+1.74}$	&	$0.883_{-0.042}^{+0.044}$	&	$0.5242$ & 15.040 \\
\hline %inserts single line
\end{tabular}
\end{table}